\documentclass[aps,prb,reprint,showpacs,twocolumn,superscriptaddress]{revtex4-2}

\usepackage[dvips]{graphicx}
\usepackage{dcolumn}% Align table columns on decimal point
\usepackage{bm}% bold mat/cite
\usepackage{xspace}
\usepackage{multirow}
\usepackage{natbib}
\usepackage{color}

\begin{document}

\preprint{}
\title{Shallow core levels, or how to determine the doping and $T_c$ of Bi$_2$Sr$_2$CaCu$_2$O$_{8+\delta}$ and Bi$_{2}$Sr$_2$CuO$_{6+\delta}$ without cooling}
\author{Tonica Valla}
\email{tonica.valla@dipc.org}
\affiliation{Donostia International Physics Center, 20018 Donostia-San Sebastian, Spain\\}
\affiliation{Condensed Matter Physics and Materials Science Department, Brookhaven National Laboratory, Upton, New York 11973, USA\\}
\author{Asish K. Kundu}
\affiliation{Condensed Matter Physics and Materials Science Department, Brookhaven National Laboratory, Upton, New York 11973, USA\\}
\affiliation{National Synchrotron Light Source II, Brookhaven National Laboratory, Upton, New York 11973, USA}
\author{Petar Pervan}
\affiliation{Institut za fiziku, Bijeni\v{c}ka 46, HR-10000 Zagreb, Croatia\\}
\author{Ivo Pletikosi\'{c}}
\affiliation{Condensed Matter Physics and Materials Science Department, Brookhaven National Laboratory, Upton, New York 11973, USA\\}
\affiliation{Department of Physics, Princeton University, Princeton, NJ 08544, USA}
\author{Ilya K. Drozdov}
\author{Zebin Wu}
\author{Genda D. Gu}
\affiliation{Condensed Matter Physics and Materials Science Department, Brookhaven National Laboratory, Upton, New York 11973, USA\\}

\date{\today}

\begin{abstract}
Determining the doping level in high-temperature cuprate superconductors is crucial for understanding the origin of superconductivity in these materials and for unlocking their full potential. However, accurately determining the doping level remains a significant challenge due to a complex interplay of factors and limitations in various measurement techniques. In particular, in Bi$_{2}$Sr$_2$CuO$_{6+\delta}$ and Bi$_2$Sr$_2$CaCu$_2$O$_{8+\delta}$, where the mobile carriers are introduced by non-stoichiometric oxygen $\delta$, the determination has been extremely problematic. Here, we study the doping dependence of the electronic structure of these materials in angle-resolved photoemission and find that both the doping level, $p$, and the superconducting transition temeprature, $T_c$ can be precisely determined from the binding energy of the Bi $5d$ core-levels. The measurements can be performed at room temperature, enabling the determination of $p$ and $T_c$ without cooling the samples. This should be very helpful for further studies of these materials.

\end{abstract}
\vspace{1.0cm}


\maketitle

\section{Introduction}

Cuprate superconductivity arises from the intricate interactions between the electrons within the copper-oxygen planes. Doping, or the introduction of foreign atoms that shift the effective hole (or electron) concentration away from the half-filling  is fundamental to turn the antiferromagnetic parent Mott insulator into a high temperature superconductor \cite{Bardeen1957,Bednorz1986,Scalapino1986,Anderson1987,Zaanen1989,Keimer2015,Varma2020}.
However, 38 years after the discovery, the mechanism of high-$T_c$ superconductivity is still unresolved.
One of the reasons is related to uncertainties in the doping level of studied materials. The doping level in cuprates has been notoriously difficult to measure directly and accurately \cite{Obertelli1992,Presland1991}.  No single technique provides a definitive picture of the doping level. In addition, the doping level $p$, defined as the concentration of holes (or electrons) away from the half filling, may differ from the concentration of mobile carriers contributing to transport, $p_{tr}$ \cite{Ando2004,Gorkov2006,Badoux2016}. Hall effect and thermoelectric power offer indirect estimates of mobile carriers $p_{tr}$, but the Hall coefficient and thermopower in cuprates are strongly dependent on temperature \cite{Presland1991,Obertelli1992,Ando2004,Gorkov2006,Badoux2016}. Each method has uncertainties, and the respective values vary substantially. 
In practice, $p$ is usually calculated backward from the measured $T_c$ by assuming the putative parabolic $T_c-p$ dependence that is considered universal for all the cuprates \cite{Presland1991,Obertelli1992}. However, only in a very limited number of materials such as La$_{2-x}$Sr$_x$CuO$_4$ and La$_{2-x}$Ba$_x$CuO$_4$, can the doping be approximately determined from chemical composition as $p\approx x$. These two materials however, have very different $T_c-p$ dependencies, illustrating the invalidity of the "universal" parabolic $T_c-p$ dome \cite{Moodenbaugh1988}.
In particular, in Bi$_{2}$Sr$_2$CuO$_{6+\delta}$ (Bi2201) and Bi$_2$Sr$_2$CaCu$_2$O$_{8+\delta}$ (Bi2212), in which the doping is induced by introduction of non-stoichiometric oxygen $\delta$, the determination has been very difficult \cite{Presland1991,Obertelli1992}. Bi2212 and Bi2201 have been prototypical cuprate high-Tc superconductors (HTSC), with their phase diagrams  heavily studied by many different techniques. In particular, these materials have been a perfect subject of angle resolved photoemission spectroscopy (ARPES) and scanning tunneling microscopy (STM) studies due to their ease of cleaving. ARPES and STM have been crucial for understanding cuprate superconductivity as these two techniques have provided the invaluable information about different phenomena and their developments with doping, mostly in Bi2212. The $d$-wave symmetry of the superconducting gap \cite{Shen1993,Damascelli2003}, the normal state gap (pseudogap) \cite{Ding1996,Marshall1996,Renner1998}, the quasiparticle (QP) self-energy \cite{Valla1999,Damascelli2003,Valla2007,Valla2020a} and, more recently, the reconstruction of the Fermi surface (FS) \cite{Fujita2014,Valla2019} are few notable examples. 

Until recently, the most studied cuprate, Bi2212, could only be doped within a limited range on the overdoped side, leaving a crucially important region of the phase diagram, where $T_\mathrm{c}\rightarrow{0}$, out of reach. Only very recently, has it become possible to extend the overdoped range beyond the point at which superconductivity vanishes by annealing the \textit{in-situ} cleaved samples in ozone \cite{Drozdov2018,Valla2020a}. This has made it possible to monitor the development of electronic excitations as superconductivity weakens and finally completely disappears, allowing a better understanding of its origins \cite{Valla2020a}. 

 By modifying the doping level of the as-grown crystals, through annealing of the \textit{in-situ} cleaved samples either in vacuum, for underdoping, or in ozone, for overdoping, we were able to span the wide region of the phase diagram, ranging from strongly underdoped, to strongly overdoped, where the superconductivity was completely suppressed. By measuring the volume of the underlying Fermi surface, we were able to precisely determine the doping level directly from ARPES. That allowed us to follow the development of spectral features with doping with unprecedented clarity and construct the phase diagram that correctly maps different phases and properties of both Bi$_2$Sr$_2$CaCu$_2$O$_{8+\delta}$ and Bi$_{1.8}$Pb$_{0.4}$Sr$_2$CuO$_{6+\delta}$. As an interesting result, we have discovered that the binding energy of the Bi {\it5d} core levels may serve as a precise indicator of doping levels and $T_c$ in both materials. As the measurements of Bi {\it5d} core levels can be performed very quickly and at room temperature, they can provide extremely efficient and precise determination of both quantities, without the need for cooling the samples.

\section{Materials and Methods}

The experiments were done in an experimental facility that integrates oxide-MBE with ARPES and SI-STM capabilities within the common vacuum system \cite{Drozdov2018,Valla2020a,Kim2022}. The starting samples were synthesized by the traveling floating zone method and the superconducting transition temperature ($T_c$) was measured by spin susceptibility down to 1.8 K. 
The \textit{as-grown} Bi2201 samples were not superconducting and they required a vacuum annealing to achieve superconductivity with the maximal $T_c\approx20-25$ K. 
The \textit{as-grown} Bi2212 samples were slightly overdoped ($T_c=91$ K) single-crystals. The samples were clamped to the sample holder and cleaved with the Kapton tape in the ARPES preparation chamber (base pressure of $3\times10^{-8}$ Pa). Thus, the use of Ag-epoxy was completely eliminated, resulting in perfectly flat cleaved surfaces and unaltered doping level (no need for epoxy curing at elevated temperatures). The cleaved samples were then annealed \textit{in-situ}  in the ARPES preparation chamber to different temperatures ranging from 450 to 950 K for several hours, resulting in the loss of oxygen and underdoping. For overdoping, the cleaved \textit{as-grown} samples were transfered to the MBE chamber (base pressure of $8\times10^{-8}$ Pa) where they were annealed in $3\times10^{-3}$ Pa of distilled O$_3$ at $650-750$ K for 1 hour. After the annealing, samples were cooled to room temperature in the ozone atmosphere and transfered to the ARPES chamber (base pressure of $8\times10^{-9}$ Pa).
Annealing of as grown crystals in O$_3$ resulted in increased doping in the near-surface region, as evidenced by the increased hole Fermi surface, reduced spectral gap and its closing temperature. The most of the crystal\rq{}s volume remained near the optimal doping upon ozone annealing. The thickness of the overdoped surface layer was in the sub-micron range, as only the thinnest, semi-transparent re-cleaved flakes showed the significant reduction in $T_c$ in susceptibility measurements. 
Therefore, the only measure of $T_c$ in the overdoped regime was spectroscopic: the temperature induced changes in the quasiparticle peak intensity, as well as the leading edge position clearly indicate $T_c$ \cite{Damascelli2003,Kondo2015,Valla2020a}. 
\textit{In-situ} annealing in vacuum resulted in homogeneous samples and the surface $T_c$ measured by ARPES showed no variation with repeated re-cleaving of the annealed crystal and was in a good agreement with the bulk susceptibility measurements.  
Our estimate of $T_c$ of the overdoped surfaces was within $\pm5$ K, except for the samples falling outside of the superconducting dome, for which the estimate was limited by the base temperature that could be reached with our cryostat (10 K).

The ARPES experiments were carried out on a Scienta SES-R4000 electron spectrometer with the monochromatized HeI (21.22 eV) radiation (VUV-5k). The total instrumental energy resolution was $\sim$ 4 meV. Angular resolution was better than $\sim 0.15^{\circ}$ and $0.4^{\circ}$ along and perpendicular to the slit of the analyzer, respectively. 
The Bi 5$d$ core level spectra were obtained by using the He II (40.8 eV) radiation at the total instrumental resolution of 20 meV.

\section{Results}

%######################################
\begin{figure*}[htpb]
\begin{center}
\includegraphics[width=14cm]{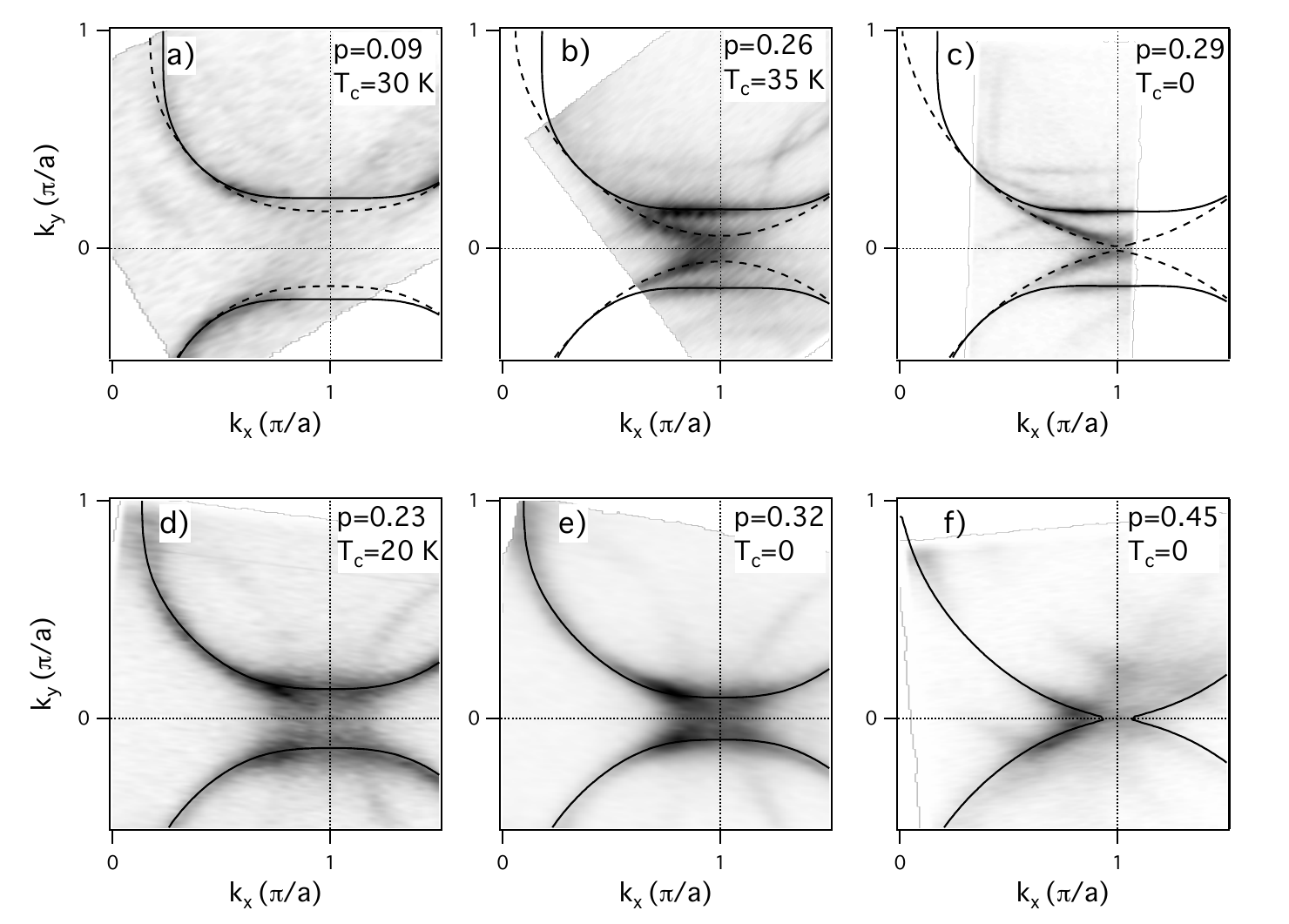}
\caption{Development of the Fermi surface of Bi$_2$Sr$_2$CaCu$_2$O$_{8+\delta}$ (Bi2212) and Bi$_{1.8}$Pb$_{0.4}$Sr$_2$CuO$_{6+\delta}$ (Bi2201) with doping. (a-c) Fermi surface of Bi2212 at various stages of doping, as indicated. (d-f) Fermi surface of Bi2201 for vacuum annealed (d), as grown (e), and as-grown, after annealing in ozone (f). Spectral intensity, represented by the gray-scale contours, is integrated within $\pm3$ meV around the Fermi level. Solid and dashed curves represent the Fermi surfaces obtained from tight-binding (TB) approximation that best fit the experimental data. 
The area enclosed by the TB lines that best represent the experimental data is calculated and used for determination of the doping parameter $p$. The maps were recorded at 12 K, for Bi2212, and at 22 K (normal state) for Bi2201, respectively.
}
\label{Fig1}
\end{center}
\end{figure*}
%######################################

Figure \ref{Fig1} shows the photoemission intensity from the narrow energy window around the Fermi level ($\pm$3 meV), representing the Fermi surfaces of the Bi2212 (a-c) and Bi2201 (d-f) samples, at different doping levels.
The tight binding (TB) contours, representing the best fits to the experimental Fermi surfaces are also shown \cite{Markiewicz2005}. The bare in-plane band structure is approximated by the tight-binding formula:

$E_{\pm}(k) =\mu - 2t (\cos k_x + \cos k_y) + 4t' \cos k_x \cos k_y - 2t'' (\cos 2k_x + \cos 2k_y) \pm t_{\perp} (\cos k_x - \cos k_y)^2 /4$

For Bi2212, $t_{\perp}\neq0$ and $\pm$ is for anti-bonding (bonding) state. $\mu$ is chemical potential. The hopping parameters that best describe the Fermi surfaces of selected measured samples are given in Table \ref{tab}.

The Bi2201 has a single Fermi sheet, while the Bi2212 has two, the bonding (marked by the solid TB contour) and the antibonding one (dashed TB contour). The \lq\lq{}shadow\rq\rq{} FS, shifted by $(\pi, \pi)$ relative to the intrinsic one is also visible. In addition, in the Bi2212 and in the ozone annealed Bi2201, the supermodulation replicas are also visible. Both structures originate from the structural supermodulations and are shown to minimally affect the electrons in the Cu-O planes \cite{Norman1995,Ding1996b,Mans2006,Valla2019}. 

%######################################
\begin{figure*}[htpb]
\begin{center}
\includegraphics[width=17cm]{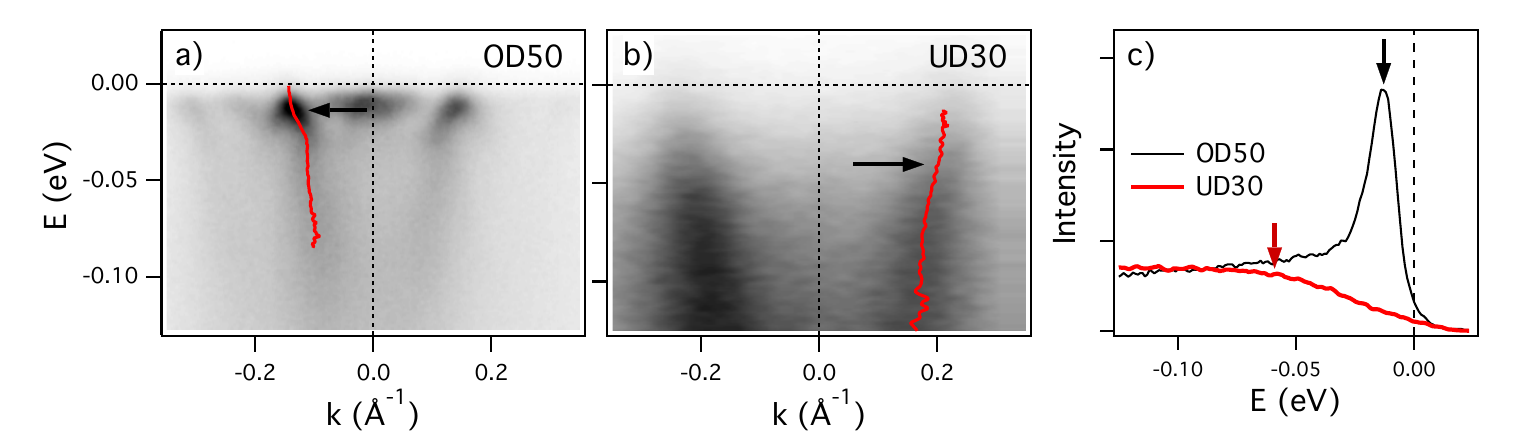}
\caption{Determination of the Fermi wave-vector for gaped states. (a) Dispersion at $k_x = \pi/a$ along the Y-M-X line corresponding to the Bi2212 sample with $T_c=50$ K $(p=0.25)$. (b) The same for the sample with $T_c=30$ K $(p=0.09)$. The red lines represent the peak position in the  momentum distribution curves (MDC) fitted by Lorentzians. The arrows indicate positions were the MDC derived dispersion starts becoming "vertical". (c) The energy distribution curves (EDC) for the two samples from (a) and (b), corresponding to the wave vector at which dispersion saturates vertically. The arrows mark the corresponding gap magnitudes, $\Delta_0$, from Drozdov \textit{et al} \cite{Drozdov2018}.
}
\label{Fig2}
\end{center}
\end{figure*}
%######################################

%######################################
\begin{figure}[htbp]
\begin{center}
\includegraphics[width=8.5cm]{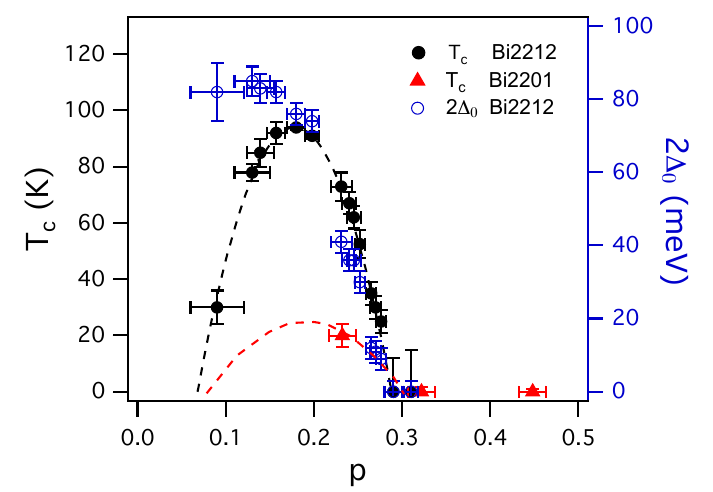}
\caption{Phase diagram of Bi2212 and Bi2201 . Superconducting transition temperature, $T_c$, for Bi2212 (black circles) and Bi2201 (red triangles) plotted versus experimentally determined doping, $p$. The maximal gap magnitude, $\Delta_0$,
for Bi2212 (blue circles) is plotted on the right axis. The experimentally determined Bi2212 superconducting dome \cite{Drozdov2018} is represented by black dashed line. The superconducting dome for Bi2201 is represented by the red dashed line \cite{Valla2021}.}
\label{Fig3}
\end{center}
\end{figure}
%######################################

For each sample, the number of carriers is obtained directly from the Luttinger count of the area enclosed  by the Fermi contour, $p=2A_{FS}-1$. For the Bi2212, both the bonding and the antibonding states are counted, $A_{FS}=(A_{-}+A_{+})/2$, originating from two Cu-O planes per unit cell. 

The Fermi wave-vectors, $k_F$, for Bi2212 were obtained at the base temperature, $T\sim10-15$ K, in the superconducting state for most of the samples. In the samples that displayed the sharp Bogoliubov's quasiparticles (BQP), and those were all the superconducting samples except for the most underdoped one, $p=0.09$, the Fermi points were extracted from energy distirbution curves (EDC), at the position where the BQP had a maximal energy (Fig. \ref{Fig2}(a,c)). In the heavily underdoped sample ($p=0.09$), the BQPs were not observed and the EDCs were monotonic functions of energy, lacking any peak-like structure at low energies (red curve in Fig. \ref{Fig2}(c)). 
We note that both in the case when the BQPs were present and when they were absent, the peaks in momentum distribution curves (MDC) at “sub-gap” energies stopped dispersing, forming a vertical dispersion of the residual intensity inside the gap (the peak was at constant $k$ as $E$ was varied inside the gap), as shown in Fig. \ref{Fig2}(a, b), illustrating two different doping cases. In the samples with BQPs present, that $k$ position was found to be equal to the position of back-folding point in Bogoliubov’s dispersion, or to the true $k_F$ point from the normal state. 
In the strongly underdoped samples where BQP did not exist, the $k$ point where the vertical dispersion is observed is taken to be the $k_F$.

Regarding the precision to which the doping level can be experimentally determined from the Luttinger count, we note that the states in the antinodal region of the Brillouin zone are much broader in the strongly underdoped samples than those in the overdoped ones, as illustrated in Fig.\ref{Fig2}. The broadening might be due to the inhomogeneity in doping and/or the increased scattering (self-energy) in strongly underdoped samples. Additional factor is the bilayer splitting, which was not clearly resolved in the Bi2212 at $p=0.09$. Regardless of the broadening mechanism, for the broader MDC peaks, the peak positions could be determined with less certainty, necessarily leading to larger uncertainties in the doping level. This was already reflected in the larger $x-$axis error bars in our previous publications \cite{Drozdov2018,Valla2020,Kundu2020}, where we estimated that $\Delta p/p\approx2\Delta k_F/k_F$. Here, $\Delta p$ and $\Delta k_F$ could be treated as uncertainties originating from standard deviations in the peak positions in fitting of MDC curves. 
However, if the MDC peaks are broad due to an intrinsic inhomogeneities (in doping, for example) in these heavily underdoped samples, then $\Delta k_F$ should be regarded as originating from a distribution of such inhomogeneities. Consequently, the width of the MDC peaks, and not only uncertainty in their positions should be taken into account when determining an uncertainty in the doping level. Although we believe that the inhomogeneities do not play the dominant role, as will be discussed later, we acknowledge that the uncertainty in the MDC peak position is not the only limiting factor in determining the doping level of a strongly underdoped sample. Therefore, our $x-$axis error bars in Fig. \ref{Fig2} and Fig. \ref{Fig5} now reflect the momentum width of the MDC peaks averaged over the Fermi surface and not uncertainties in the peak positions.

\begin{table}
\caption{\label{tab}Tight-binding parameters for Bi2201 and Bi2212 samples from Fig.\ref{Fig1}}
\begin{ruledtabular}
\begin{tabular}{lccccc}
Sample & $\mu$ (eV)& $t$ (eV)& $t'$ (eV)& $t''$ (eV)& $t_{\perp}$ (eV)\\
\tableline
Bi2212(a) & 0.34 & 0.39 & 0.12 & 0.045 & 0.108\\
Bi2212(b) & 0.445 & 0.36 & 0.108 & 0.036 & 0.108\\
Bi2212(c) & 0.467 & 0.36 & 0.108 & 0.036 & 0.108\\
Bi2201(d) & 0.42 & 0.36 & 0.1 & 0.036 & 0\\
Bi2201(e) & 0.49 & 0.36 & 0.1 & 0.036 & 0 \\
Bi2201(f) & 0.545 & 0.37 & 0.1 & 0.036 & 0\\
\end{tabular}
\end{ruledtabular}
\end{table}

%%#####################################
\begin{figure}[ht]
\begin{center}
\includegraphics[width=7cm]{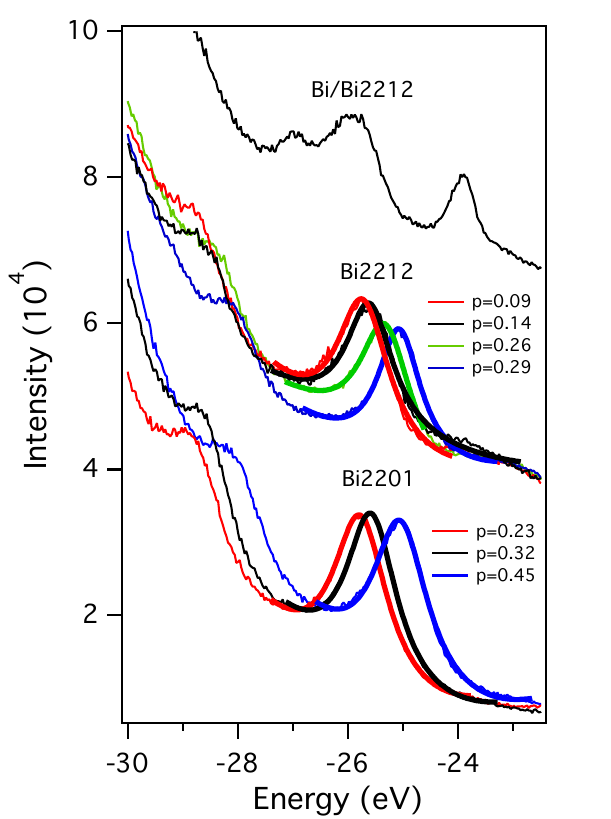}
\caption{Doping dependence of Bi {\it5d} core levels in Bi2212 and Bi2201. The Bi {\it5d} levels spectra for several Bi2212 (middle) and Bi2201 (bottom) samples at different doping levels, as indicated. The top spectrum corresponds to a 1.5 \AA\ thick film of Bi on \textit{as grown} Bi2212. The presented spectra were obtained by using the He II radiation at $12-20$ K. Spectra were also recorded at 300 K, with no changes observed with temperature. The thick lines represent the fits of the Bi 5$d_{5/2}$ core level with Lorentzian peak on a cubic background. The spectra for Bi2212 and Bi/Bi2212 are offset in $y$-direction for clarity.  
}
\label{Fig4}
\end{center}
\end{figure}
%######################################

By having a direct measure of the doping level, we can now plot various other quantities versus independently determined $p$ and construct our own doping phase diagrams. For example, we can plot $T_c$ measured by ARPES (as described in Drozdov \textit{et al} \cite{Drozdov2018}), or by the bulk susceptibility measurements, versus doping, as shown in Fig. \ref{Fig3} for both Bi2212 and Bi2201. We show the doping phase diagram with the superconducting phases for both Bi2212 from Drozdov \textit{et al} \cite{Drozdov2018} and the Bi2201 from Valla \textit{et al} \cite{Valla2021}. We also show the maximal gap magnitude, $\Delta_0$, as a function of doping for Bi2212. These phase diagrams generally agree with the published ones for these compounds \cite{Presland1991,Obertelli1992,Damascelli2003,Keimer2015,Ding2019}. The only adjustment needed for Bi2212 was the slight shift of the SC dome, now centered at $p = 0.18$. We also note that we have now revised the $\Delta_0$ value for the $p=0.09$ sample. As can be seen in Fig. \ref{Fig2}(c), the EDC from that sample lacks any discernible peaks and the gap cannot be determined in a conventional way, by associating it with either the peak or the leading edge energy \cite{Tanaka2006}. In that case, we think that a much better estimate for the gap magnitude is the energy at which the MDC-derived dispersion starts turning vertical, as illustrated in Fig. \ref{Fig2}. In cases where the BQPs do exist, that point coincides with the maximum in the Bogoliubov dispersion and the associated peak energy of the corresponding EDC. In the $p=0.09$ sample, there is no peak in EDC, but the upturn in the MDC-derived dispersion exists, indicating the near-vertical dispersion of in-gap intensity. We identify this point as a much better estimate for the gap magnitude in the cuprate superconductors without BQPs. In fact, we believe that this method might be appropriate for any gaped system with no sharp quasiparticle peaks.  

Also shown in Fig. \ref{Fig3} is the phase diagram for Bi2201. However, as Bi2201 is typically co-doped with Pb or La, its phase diagram is not unique and comparisons should be done only for the same compounds.

Now we want to point to another quantity of Bi2212 and Bi2201 spectra, seemingly unrelated to superconductivity, but which displays the doping dependence that can be used to directly and quickly determine the doping level and $T_c$ in these materials, without even performing any $T$-dependent measurements. This quantity is the binding energy of the Bi $5d$ core-level states inside the Bi-O planes. 

Figure \ref{Fig4}(a) shows the Bi 5$d$ region of the photoemission spectra corresponding to several Bi2212 and Bi2201 samples at different doping levels. We see that with the increase in doping, the states shift to lower binding energies, both in Bi2212 and Bi2201. The Bi 5$d$ and Sr 4$p$ levels both shift when the doping is varied, albeit by different amounts. They both move to lower binding energies with doping. This indicates that both Bi and Sr are always completely oxidized, with a stable valence, in their corresponding planes and that they cannot be more oxidized. Then, with more oxygen content in the sample upon ozone annealing, the Bi and Sr core levels shift to lower binding energies due to a better screening of the photohole in an increasingly metallic environment of the progressively more doped Cu$-$O planes. This is consistent with conventional description of core-hole screening in metals \cite{Johansson1980}.

It is interesting to note that the width of the Bi 5$d$ states is nearly the same for all the samples. If the doping was very inhomogeneous in the heavily underdoped samples, one would expect that the Bi 5$d$ peaks should also become broader. Therefore, we do not think that the doping inhomogeneity is as severe as some scanning tunneling microscopy studies suggest \cite{Pan2001}. Consequently, the inhomogeneity cannot be entirely responsible for the observed broadening of the Fermi surface in the antinodal region in strongly underdoped samples. 

We note that the observed relationship between the doping and the binding energy of Bi $5d$ states works regardless of the doping mechanism. In particular, we have discovered that any metallic species adsorbed on the surface of Bi2212 will donate electrons to it and consequently will reduce the hole doping at the interface \cite{Kundu2021,Kundu2020}. The hole doping can be reduced so severely that the films on Bi2212 often make the interface non-superconducting, rendering the proximity effect at the metal - HTSC interfaces impossible \cite{Yilmaz2014,Kundu2020,Kundu2021}.

For an illustration, we also show a spectrum of the thin Bi film deposited on Bi2212 that was near the optimal doping ($p=0.2$, $T_c\approx 91$ K). The Bi $5d_{5/2}$ level corresponding to the Bi$-$O plane was shifted to $-25.86$ eV, while the state corresponding to the film was at $-23.87$ eV. The Bi film itself was not connected and the bands were not formed yet. Its main effect on the Bi2212 samples was donating electrons to the Cu$-$O planes.

The shifts in the Bi $5d_{5/2}$ level are summarized in Fig. \ref{Fig5}. The first two points for Bi2212 in Fig. \ref{Fig5}(a) ($p\approx0$ and $p\approx0.045$) correspond to the doping levels after depositing Bi films on an optimally doped Bi2212 sample, clearly indicating severe underdoping, outside the superconducting phase. The dependence on the doping level for both Bi2212 (blue symbols) and Bi2201 (green symbols) indicate that in both materials the Bi $5d$ levels shift closer to the Fermi level nearly linearly with increasing doping (the parameters for some of the Bi2212 samples used in this study are given in Table \ref{tab1}). One significant distinction between Bi2201 and Bi2212 is that there is a $\sim0.5$ eV difference in the Bi 5$d_{5/2}$ level energy at similar doping levels. This points to a higher metallicity and a better screening of charges in Bi2212 than in Bi2201 - not unexpected, as the two Cu$-$O similarly doped planes should certainly screen better than a single one. 

%%#####################################
\begin{figure}[htbp]
\begin{center}
\includegraphics[width=8cm]{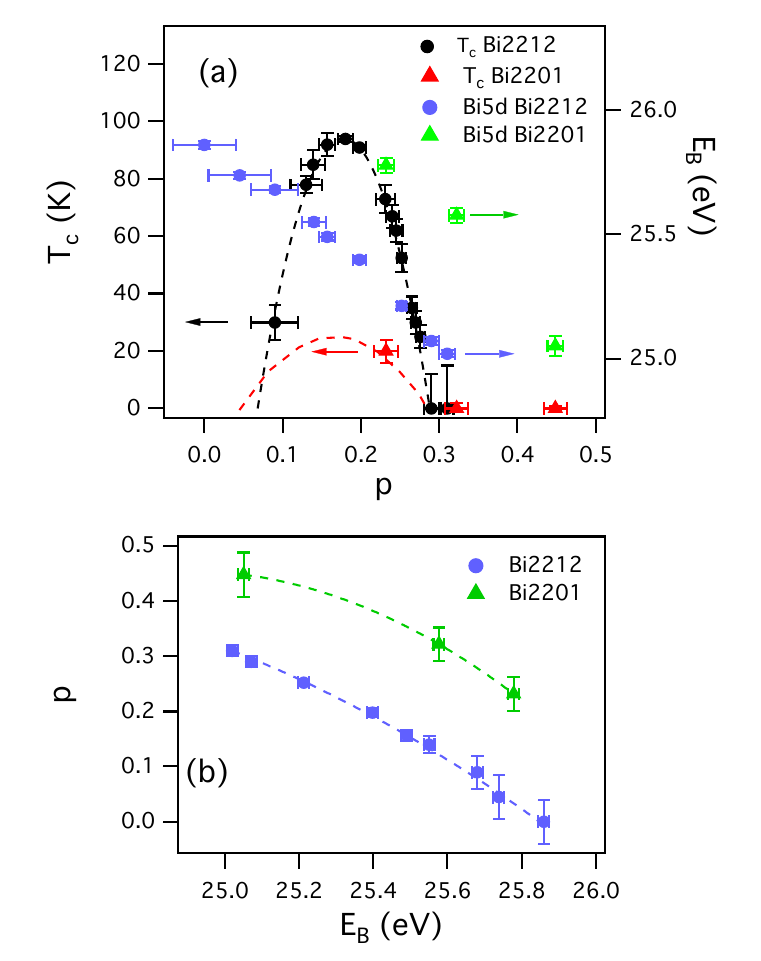}
\caption{The phase diagrams of Bi2212 and Bi2201. (a) The binding energy of Bi $5d_{5/2}$ level in Bi2212 (blue circles) and Bi2201 (green triangles)  samples as a function of doping on the right-hand scale. $T_c$ is also shown on the left-hand scale. (b) The doping level, $p$ as a function of binding energy of Bi $5d_{5/2}$ core level in Bi2212 and Bi2201. Dashed lines represent parabolic fits of the data.
}
\label{Fig5}
\end{center}
\end{figure}
%######################################
Figure \ref{Fig5} (b)  shows the inverse dependence, $p$ vs $E_B(5d_{5/2})$, the main result of this study that allows the direct determination of the doping level of Bi2212 and Bi2201 samples by measuring solely the spectra of Bi $5d$ core levels. In the studied doping regime, the parabolic fit can be used to determine the doping level from the binding energy $E_B$ of the Bi $5d_{5/2}$ level: 
\begin{equation}
    p=a+bE_B+c(E_B)^2
\end{equation}
where $a=-77.468$, $b=6.4799$ eV$^{-1}$ and $c=-0.13474$ eV$^{-2}$ for Bi2212 and $a=-177.06$, $b=14.261$ eV$^{-1}$ and $c=-0.28641$ eV$^{-2}$ for Bi2201.

\begin{table}
\caption{The parameters of several Bi2212 samples from Fig.\ref{Fig5}\label{tab1}}
\begin{ruledtabular}
\begin{tabular}{lccc}
$p$	& $T_c$ (K) & treatment & Bi $5d_{5/2}$ (eV))\\
\tableline
0 & NA & Bi film on as grown & 25.86 \\
0.09 & 30 & vacuum ann. (950 K) & 25.70 \\
0.14 & 85 & vacuum ann. (800 K) & 25.55 \\
0.16 & 92 & vacuum ann. (450 K) & 25.49 \\
0.20 & 91 & as grown & 25.40 \\
0.25 & 50 & O$_3$ annealed & 25.21 \\
0.29 & 0 & O$_3$ annealed & 25.05 \\
0.31 & 0 & O$_3$ annealed & 25.02 \\
\end{tabular}
\end{ruledtabular}
\end{table}

By having the doping level determined, one can then use the "universal" formula to determine $T_c$ of an unidentified sample \cite{Presland1991}:
\begin{equation}
    T_c=T_c^{max}[1-82.6(p-p^{max})^2]
\end{equation}
where $T_c^{max}=94$ K, and $p^{max}=0.18$ for our modified Bi2212 dome. The parameters for Bi2201 depend on Pb or La co-doping and were not precisely determined for our Bi$_{1.8}$Pb$_{0.4}$Sr$_2$CuO$_{6+\delta}$ sample.

\section{Discussion}

As demonstrated in Fig. \ref{Fig5}, our study enables an easy determination of $p$, a quantity previously notoriously difficult to determine, and $T_c$, without even cooling the sample. The determination can be performed in the \textit{in-lab} environment, using an UV-lamp and He II excitation, as in our study. The measurements can be performed very quickly, due to a large cross section for photoemission of Bi $5d$ levels at He II excitation. However, the other excitation sources (synchrotron or \textit{in-lab} x-ray sources) and different Bi (or Sr) core levels could be used for the determination. For different core levels, the slope of the doping dependence might be slightly different, but the overall shape (nearly linear) should be preserved. In that case, re-calibration of the binding energy vs $p$ relationship could be performed with just two different samples with known $T_c$. We also expect that other core levels with stable oxidation states could serve as indicators of the doping level of Cu-O planes in other cuprates. Their shift to lower binding energies would indicate better screening due to higher metallicity of the neighbouring Cu-O planes as the hole doping increases. We hope that this new method will be very helpful in constructing precise phase diagrams in future studies of cuprate superconductors. 

Finally, we note that some earlier studies where the doping was varied not by oxygen non-stoichiometry, but by substituting Ca by Y, display somewhat different behavior \cite{Tjernberg1997}. Our present study cannot be directly compared, as we have not studied how the hole doping and core levels depend on Ca$-$Y, Bi$-$Pb and other non-oxygen non-stoichiometries. Unfortunately, such studies have never been done systematically. We think that the main problem with Tjernberg \textit{et al} \cite{Tjernberg1997} was a lack of an access to the actual doping level of Cu$-$O planes. The shifts were presented vs. yttrium content that does not equate the doping $p$.

In some other studies, the doping levels were deduced from the Luttinger count, as required, but the observed core levels shifts were interpreted as rigid shifts due to variation of chemical potential $\mu$ during doping \cite{Hashimoto2008}. Even though there were obvious inconsistencies with that scenario, as the different core levels shifted differently, that fact was ignored for a long time. Our present study clearly demonstrates that the shift in chemical potential is not the main reason of the observed core levels shifts.
Our study provides an independent and consistent estimate of the chemical potential $\mu$ and of its dependence on the measured doping level from the tight binding approximation for the valence band, as shown in Table 1 for selected samples. If we take this as a realistic estimate, it is obvious that $\mu$ shifts roughly 4-5 times less than the studied Bi $5d$ core levels over the same doping range and cannot be responsible for the core level shifts observed here. On the other hand, if the increased screening due to higher metallicity of the doped Cu$-$O planes is the main  driver of the observed shifts, then we would expect that as the material turns from $p-$type into $n-$type (by depositing a metallic film on an underdoped sample, for example), the Bi and Sr core levels should eventually start shifting back, near $p=0$. Previous studies on different families of $n-$type cuprates were inconclusive and contradicting \cite{Harima2001,Vasquez2001}, indicating that further studies are needed. Again, the precise determination of $p$ (via the Luttinger count) and the binding energies of the core levels should be performed simultaneously. We note that our least doped point $(p\approx0)$, where the underdoping was achieved by Bi deposition, suggests that it should be possible to turn the surface layers of Bi2212 into $n-$type and resolve this issue in future studies.

Irrespectively of the origin of the observed shifts, we have discovered a simple and efficient method of determining the doping level of the Cu-O planes and $T_c$ of Bi-based cuprate superconductors from purely spectroscopic measurements that does not require cooling.

\section{acknowledgments}
This work was supported by the US Department of Energy, Office of Basic Energy Sciences, contract no. DE-SC0012704. I.P. was supported by  ARO MURI program, grant W911NF-12-1-0461. T.V. acknowledges the support from the Red guipuzcoana de Ciencia, Tecnología e Innovación – Gipuzkoa NEXT 2023 from the Gipuzkoa Provincial Council.

%\bibliography{Cuprates}
%

\end{document}